\documentclass[fleqn,usenatbib,preprint, twocolumn]{mnras}

\usepackage{newtxtext,newtxmath}

\usepackage[T1]{fontenc}
\usepackage{ae,aecompl}

\usepackage{graphicx}	
\usepackage{amsmath}	
\usepackage{amssymb}	
\usepackage[usenames, dvipsnames]{color} 
\usepackage{float}
\usepackage{subcaption}

\title
[UHECRs from radio galaxies]
{Fornax A, Centaurus A 
and other radio galaxies as sources of 
ultra-high energy cosmic rays}

\author[J.~H.~Matthews et al.]{J.~H.~Matthews$^1$\thanks{james.matthews@physics.ox.ac.uk}, A.~R.~Bell$^2$, K.~M.~Blundell$^1$
and A.~T.~Araudo$^3$
\\$^1$University of Oxford, Astrophysics, Keble Road, Oxford, OX1 3RH, UK
\\$^2$University of Oxford, Clarendon Laboratory, Parks Road, Oxford OX1 3PU, UK
\\$^3$Astronomical Institute, Czech Academy of Sciences, Bo\v{c}ni II 1401, CZ-141 00 Prague, Czech Republic
}

\date{Accepted to MNRAS Letters 2018 May 31. Received 2018 May 30; in original form 2018 April 26.}
\pubyear{2018}

\begin{document}
\label{firstpage}
\pagerange{\pageref{firstpage}--\pageref{lastpage}}
\maketitle

\begin{abstract}
The origin of ultra-high energy cosmic rays (UHECRs)
is still unknown. It has recently been proposed that 
UHECR anisotropies can be attributed to 
starburst galaxies or active galactic nuclei. 
We suggest that the latter is more likely 
and that giant-lobed radio galaxies such as Centaurus A and 
Fornax A can explain the data.
\end{abstract}




\section{Introduction}
Although their origin is unclear, ultra-high energy cosmic rays
(UHECRs) have often been posited to come from radio galaxies, the subclass of 
active galactic nuclei (AGN) that are luminous at radio frequencies. 
One fundamental reason for this
is that radio galaxies allow the \cite{hillas_origin_1984} criterion 
to be met; they are large, produce fast, energetic outflows, 
and have reasonably high magnetic fields. 
They are also known to produce high-energy electrons \citep[e.g.][]{hargrave_observations_1974,croston_high-energy_2009}, 
thought to be mostly accelerated 
by diffusive shock acceleration \citep[DSA;][]{axford_acceleration_1977,krymskii_regular_1977,blandford_particle_1978,bell_acceleration_1978} at termination shocks, which create the
hotspots seen in \cite{fanaroff_morphology_1974} (FR) type II sources.

Beyond this general physical reasoning, observational results from CR observatories have also hinted at an association between AGN and UHECRs. Initial results from the Pierre Auger Observatory (PAO) indicated a tantalising correlation between UHECR arrival directions and AGN source catalogues \citep{pierre_auger_collaboration_correlation_2007}. However, this correlation declined in significance as more data were obtained \citep{abreu_update_2010} and subsequent studies found only low-significance departures from isotropy \citep[e.g.][]{pierre_auger_collaboration_search_2012}.

Recently, results from the PAO
indicated a large-scale anisotropy in the arrival directions 
of UHECRs \citep{pierre_auger_collaboration_observation_2017}. 
Departures from anisotropy were also confirmed 
on intermediate angular scales \citep[][hereafter A18]{the_pierre_auger_collaboration_indication_2018}, leading to a 4$\sigma$ association with starburst galaxies (SBGs) and a slightly weaker association with $\gamma$-ray AGN from the 2nd catalogue of hard {\it Fermi}-LAT sources \citep[2FHL;][]{ackermann_2fhl:_2016}. The radio galaxy
Fornax A does not appear in the 2FHL catalogue. 
In this Letter, we show that including Fornax A in the analysis could explain the observed excess at southern Galactic latitudes, which could 
increase the significance of the 
$\gamma$-AGN association. We also outline the physical reasoning behind this and discuss parallels with Centaurus A and other sources. In addition, we show that the minimum power needed to accelerate protons up to 10~EeV can be supplied by jet outbursts in radio galaxies but not by starburst winds.

\vspace{-2em}
\section{UHECR Arrival Directions}
A18 analysed the PAO dataset 
consisting of 5514 events above 20 EeV, finding that 
a number of models can provide a better fit than isotropy. In particular,
a model involving SBGs is favoured over
isotropy at the 4$\sigma$ level, while their
alternative models involving
AGN attain lower significance (2.7$\sigma$-3.2$\sigma$).
The threshold energy above which the correlation
is evaluated is scanned by A18 
to find the best value and the relevant statistical penalty is taken into account when evaluating
the above significance levels. 
A18 find threshold energies of 
$39$~EeV for SBGs and $60$~EeV for $\gamma$-AGN. 

\begin{figure*}
\centering
\includegraphics[width=0.75\linewidth]{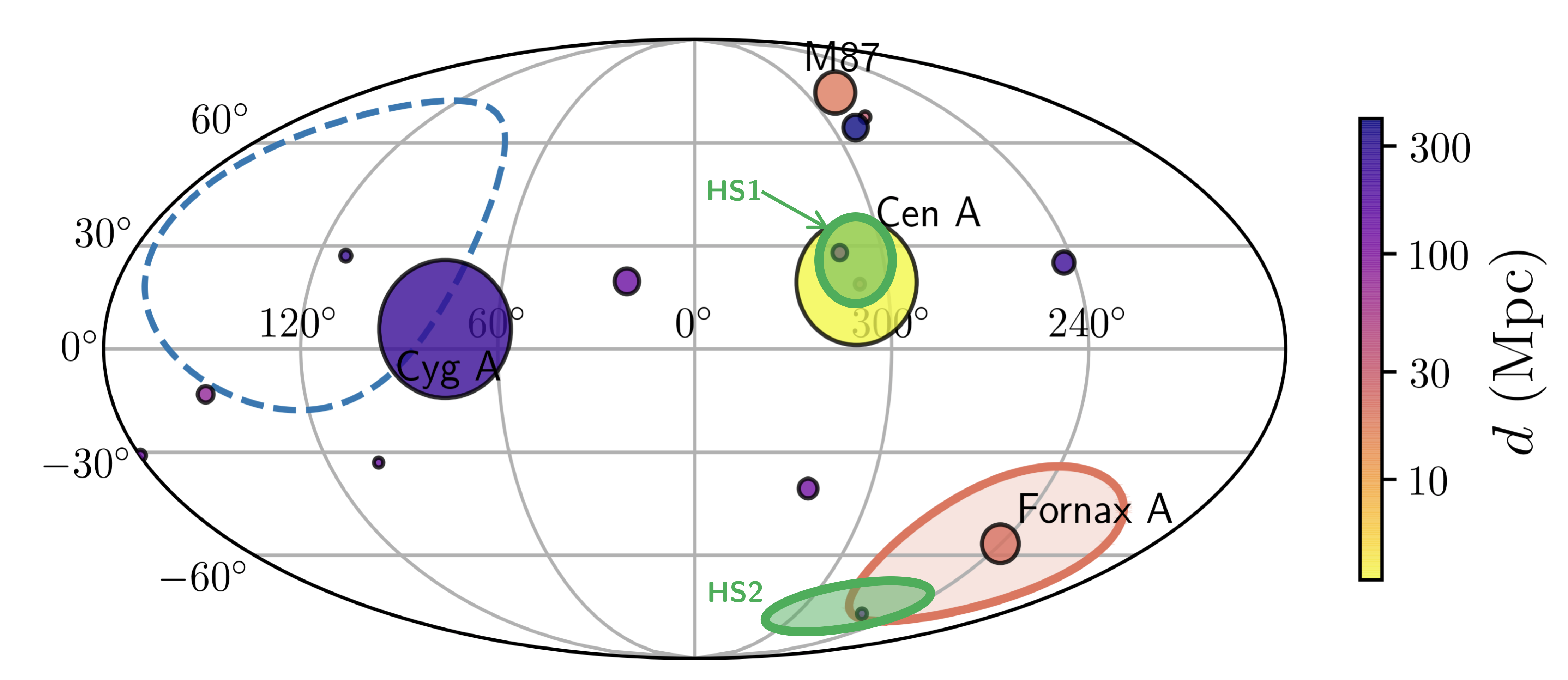}
\caption{The positions of the 16 brightest 
radio galaxies in Galactic coordinates,
with the area of the points proportional 
to 1.1 GHz radio flux and colour corresponding
to distance from Earth. The radio flux is calculated
from table 2 of vV12. The orange circle around Fornax A 
illustrates a deflection angle of $22.5^\circ$, while the 
green shaded regions mark the approximate PAO excesses above 60 EeV
(HS1 and HS2) from A18 as described in the text. The
blue dashed line marks the area of the sky inaccessible to
PAO. The projection is the same as that of fig.~7 of A18, with 
image coordinates $(x,y)$ mapped to Galactic coordinates in 
degrees $(l,b)$ by $x=\lambda \cos \theta$, $y=b$ where 
$\sin \theta= b / 90^\circ$ and $\lambda = -l$ (for $l\leq180$), $\lambda = 360^\circ - l$ (for $l\geq180$). 
}
\label{fig1}
\end{figure*}

The observed excess map above 60 EeV from A18 has two 
fairly clear hotspots (see their fig.~7). 
We do not have access to the A18 dataset, but
we can estimate the approximate positions of 
the hotspot centroids in Galactic coordinates as 
$(l=308^\circ,b=26^\circ)$ (HS1) and 
$(l=275^\circ,b=-75^\circ)$ (HS2). 
We show the two hotspots in Fig.~\ref{fig1} using 
the same projection as used by A18, together with the 16 brightest
radio galaxies from the \citet[][hereafter vV12]{van_velzen_radio_2012}
radio catalogue. In the A18 SBG fit, HS1 can be 
attributed to combined UHECR emission from M83 and NGC4945, 
while HS2 can be explained by NGC 253. In their
$\gamma$AGN fit, Centaurus A 
dominates the map with a small contribution from 
M87 (Virgo A); HS1 is associated with Centaurus A, 
while HS2 is unaccounted for. 

\subsection{Fornax A}
Fornax A (NGC 1316) is one of the brightest radio galaxies 
in the sky at 1.4 GHz \citep{schweizer_optical_1980}, 
with a flux density of 150 Jy \citep{brown_ubiquitous_2011} 
and at a distance of $20.9$Mpc (vV12). It has giant lobes
$\sim 300$~kpc across, which are bright in radio 
\citep{ekers_large-scale_1983,geldzahler_radio_1984},
as well as being one of the two objects whose lobes are 
high-energy $\gamma$-ray sources 
\citep{ackermann_fermi_2016} -- the other is Centaurus A \citep{abdo_fermi_2010,abdo_fermi_2010-1}.
However, Fornax A does not appear in the 2FHL catalogue
as it is an extended source with a
$0.15^\circ$ offset between the radio 
and gamma-ray position \citep{ackermann_fermi_2016}, 
although it is present in 3FHL \cite{ajello_3fhl:_2017}.
The absence from 2FHL meant that it was 
not included in the A18 analysis. 
Fornax A lies at a southern Galactic latitude, with the position 
of its radio core at $(l=240.16^\circ,b=-56.69^\circ)$ 
\citep{geldzahler_radio_1984}.
It can be seen to the lower right of Fig.~\ref{fig1}.
The angular separation between Fornax A and our estimated
HS2 position is $22.5^\circ$. 

\subsection{Magnetic Deflection}
The magnetic deflection of UHECRs depends on the magnetic field 
encountered -- its strength and topology -- and the 
{\em magnetic rigidity} of the UHECR, given by 
${\cal R} = E / Ze$, where $E$ is the 
CR energy and $Ze$ is the charge on the nucleus.
The deflection magnitude can then be written as 
$\theta_d = K / {\cal R}$, where $K$ is a constant depending on the magnetic field 
between the source and Earth. 
Using the \cite{jansson_new_2012}
model for the Galactic magnetic field, \cite{smida_ultra-high_2015} 
find $K=242^\circ$EV (degree exa-Volts) for Fornax A. 
For a nucleus of ${\cal R}=10$~EV, 
this corresponds to a deflection of 
$24.2^\circ$, very close to the  
offset between the PAO excess and Fornax A. 

The deflection angle in an extragalactic turbulent field
can also be estimated assuming some coherence 
length for the magnetic field, typically 1 Mpc. 
\cite{sigl_ultrahigh_2003,sigl_ultrahigh_2004} and 
\cite{eichmann_ultra-high-energy_2018} find deflections of 
12-24$^\circ$ are reasonable for a nucleus of ${\cal R}=10$~EV
travelling $20.9$~Mpc in a 1-2~nG magnetic field. 
The fact that Centaurus A, 
at a distance of only $3.7$~Mpc \citep{tully_two_2015} 
is offset from HS1 by $7^\circ$ again implies 
that large deflections are feasible for Fornax A at 
the greater distance of $20.9$~Mpc. 
Further detailed modelling work is possible, 
using tools such as CRPropa \citep{alves_batista_crpropa_2016};
however, for the purposes of this 
Letter we note that a deflection of $\approx20-25^\circ$ 
is highly plausible for a source at $20.9$Mpc,
as shown by \cite{sigl_ultrahigh_2004} and 
\cite{smida_ultra-high_2015}.

\begin{table}
\centering
    \begin{tabular}{l c c}
    
    \hline 
    Source &  2FHL  & 3FHL  \\
    & \multicolumn{2}{ c }{($10^{-12}$~erg~cm$^{-2}$~s$^{-1}$)} \\
    \hline 
    Cen A core & $3.90\pm2.29$ & $7.40\pm1.90$\\
    M87 & $5.12\pm3.47$ & $9.55\pm3.26$\\
    Fornax A & -- & $2.59\pm1.27$ \\
    \hline 
    \end{tabular}
  \caption{$\gamma$-ray fluxes for the three sources discussed in section~\ref{sec:fluxes}
}
  \label{tab:fluxes}
\end{table}

\subsection{Attenuation and Fluxes}
\label{sec:fluxes}

The $ \gamma $-ray fluxes of Centaurus A, M87 and Fornax A from the 2FHL and 3FHL catalogues 
are given in Table 1, obtained from \cite{ackermann_2fhl:_2016} and \citep{ajello_3fhl:_2017}.
A18 use the 2FHL $\gamma$-ray luminosity as a proxy for UHECR luminosity, with a choice of three scenarios for UHECR attenuation during propagation. This attenuation is due to the 
Greisen--Zatsepin--Kuzmin \citep[GZK;][]{greisen_end_1966,zatsepin_upper_1966} effect and
photodisintegration \citep{stecker_photodisintegration_1999}.
The starburst galaxies in their sample are nearby and the choice of attenuation scenario makes little difference.  Strong attenuation (scenario A) is favoured in their AGN analysis since (i) it accounts for the negligible UHECR signal from the direction of M87 which is at five times the distance of Centaurus A  (ii) Fornax A, at a distance of $20.9$~Mpc, is not included and might otherwise account for the HS2 hotspot if attenuation were weaker.

Given that M87 and Fornax A are at similar distances and that there is an Auger hotspot close to Fornax A but not M87, a successful model for the observed PAO anisotropy requires the attenuation to be less severe than scenario A of A18 and that M87 is intrinsically less luminous in UHECRs than Fornax A as we argue in section~\ref{sec:lobes}.  
Less severe attenuation would be consistent with results from the CRPropa code as given in fig.~1 of \cite{alves_batista_effects_2015}, as well
as canonical values of the GZK length of 50-100~Mpc \citep[e.g.][]{dermer_ultra-high-energy_2009,de_domenico_influence_2013}.
Sensitivity to composition and source energy spectrum makes the adoption of a single
attenuation length difficult; for example, protons at $10$~EeV and $100$~EeV have 
approximate GZK lengths of $1000$~Mpc and $100$~Mpc, respectively 
\citep{dermer_ultra-high-energy_2009}.
Approximate attenuation lengths for N$^{14}$ and Fe$^{56}$ 
nuclei at $100$~EeV are $6$~Mpc and $300$~Mpc, respectively 
\citep{alves_batista_effects_2015}.

The correlation with AGN in A18 would also be improved by including the contribution from the lobes of Centaurus A, which are estimated to be at least as bright as the core in $\gamma$-rays  \citep{abdo_fermi_2010}.
Furthermore, although there may be a direct relation between the observed $\gamma$-rays and UHECRs 
\citep{sahu_hadronic-origin_2012,yang_deep_2012,joshi_very_2018}, $\gamma$-ray luminosity may not be the best proxy for UHECR luminosity.

\section{Fading Radio lobes as UHECR reservoirs}
\label{sec:lobes}
As shown by \cite{waxman_cosmological_1995,waxman_high-energy_2001} 
and \cite{blandford_acceleration_2000}, 
there is a minimum power requirement for particle 
acceleration to high energy at shocks. 
This can be derived just from considering the magnetic 
energy density, $U_{\mathrm{mag}} = B^2/(2 \mu_0)$, 
and the Hillas energy $E_H=uBLZe$, where $B$ is the magnetic field
strength, $u$ is the shock velocity and $L$ is a characteristic size.
The maximum magnetic power delivered through a shock is then 
roughly $u L^2 U_{\mathrm{mag}}$, 
meaning we can write an equation for the minimum power 
needed to accelerate a nucleus to a given
rigidity, ${\cal R}$: 
\begin{equation}
P_{\mathrm{min}} = \frac{{\cal R}^2}{2 \mu_0 u}, 
\end{equation}
which is equivalent to
\begin{equation}
P_{\mathrm{min}} \sim 
10^{43}~\mathrm{erg~s}^{-1} 
~\left( \frac{u}{0.1c} \right)^{-1}
~\left( \frac{{\cal R}}{10 \,\rm EV} \right)^2.
\label{eq:power}
\end{equation}
Here we conservatively assume maximum efficiency and adopt $u=0.1c$
due to the difficulties with accelerating UHECRs at highly relativistic
shocks \citep{bell_cosmic-ray_2018}.
This equation is quite general and 
places a fundamental constraint on UHECR sources. 
We note that starburst 
winds struggle to meet this constraint as they have powers on the
order of $10^{42}$~erg~s$^{-1}$ and low 
shock velocities 
\citep[$\sim1000~$km~s$^{-1}$;][]{heckman_nature_1990,anchordoqui_unmasking_2017,romero_particle_2018}.

\begin{figure}
\centering
\includegraphics[width=1.1\linewidth]{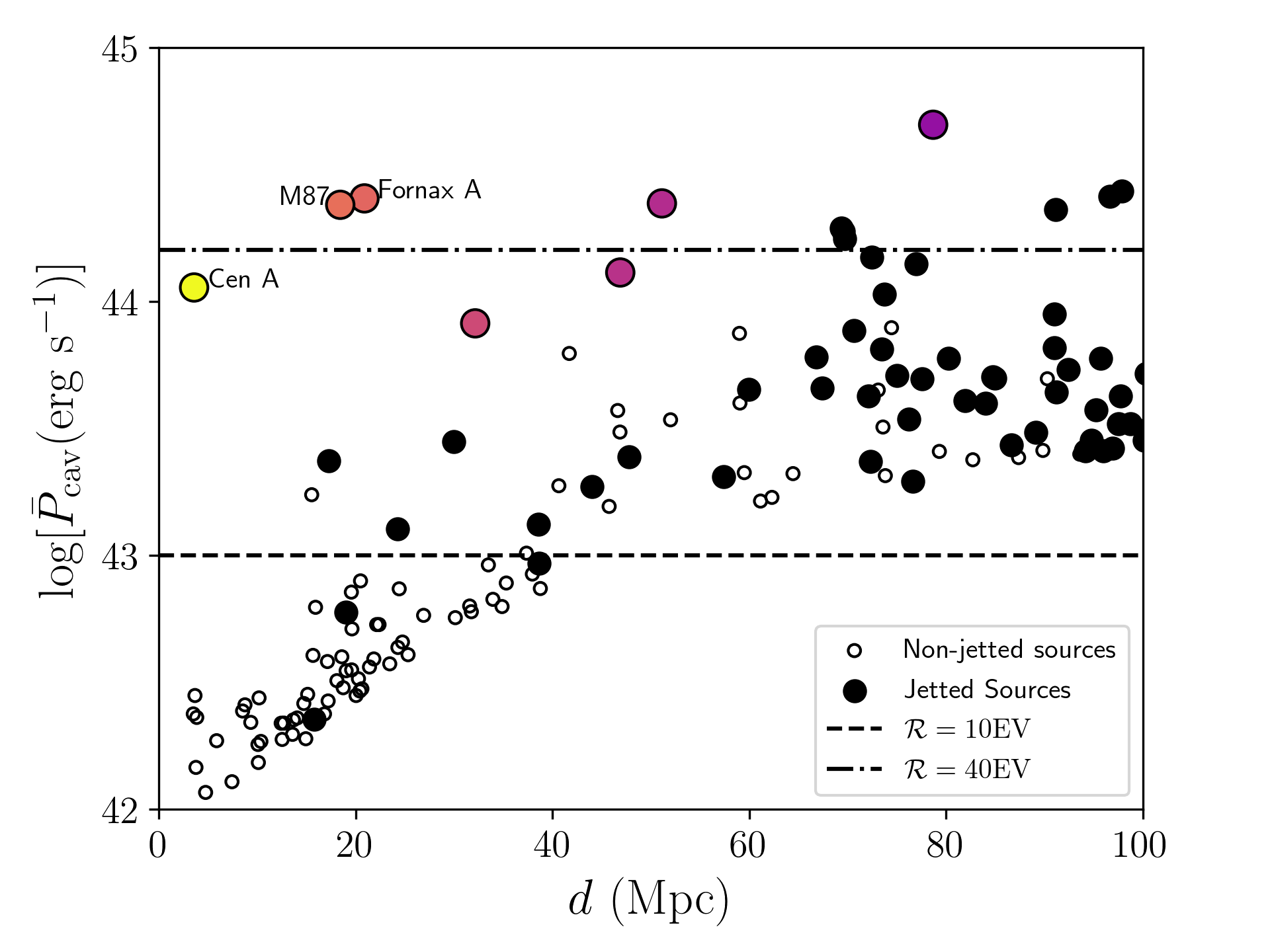}
\caption{The logarithm of estimated 
cavity power for local radio galaxies 
plotted against distance, calculated as described in section~\ref{sec:lobes}. 
The filled circles represent AGN observed to have jets and 
the coloured circles are the subset of these that are shown in
Fig.~\ref{fig1}, also with matching colours to Fig.~\ref{fig1}. 
The two horizontal lines show $P_{\mathrm{min}}$ for two different 
rigidities.}
\label{fig2}
\end{figure}

To examine which nearby radio galaxies meet the $P_\mathrm{min}$ requirement, 
we estimate a `cavity power', 
$\bar{P}_\mathrm{cav}$, using the mean empirical
relationship of \cite{cavagnolo_relationship_2010}.
This is quoted in their section~5 and given by
\begin{equation}
\bar{P}_{\mathrm{cav}} \approx 5.8 \times 10^{43} 
\left( \frac{{P}_\mathrm{radio}}{10^{40}~\mathrm{erg~s}^{-1}} \right)^{0.7}~\mathrm{erg~s}^{-1}~,
\end{equation}
where we take the 1.1 GHz luminosity from vV12 as our 
${P}_\mathrm{radio}$. This estimate should be thought
of as a rough proxy for average kinetic power, 
since we make use of the current radio
luminosity but \cite{cavagnolo_relationship_2010} relate
this to kinetic power using work done excavating a cavity.
Fig.~\ref{fig2} shows $\bar{P}_\mathrm{cav}$ 
plotted against distance,
with the power requirement from equation~\ref{eq:power} marked
for two rigidities. Cen A, Fornax A and M87 are three of only a 
handful of sources within a characteristic GZK radius 
of $50-100$~Mpc capable 
of accelerating UHECRs to ${\cal R}=10$~EV and above. 
However, the actual current jet power in these sources
is likely lower, with approximate 
estimates in the literature of 
$10^{42}$~erg~s$^{-1}$ 
\citep[Fornax A; ][]{russell_radiative_2013}
$6-8\times10^{42}$~erg~s$^{-1}$ 
\citep[M87; ][]{rafferty_feedback-regulated_2006,russell_radiative_2013} and 
$10^{43}$~erg~s$^{-1}$\citep[Cen A; ][]{russell_radiative_2013,wykes_mass_2013}.
These estimates are uncertain and rely on the enthalpy 
($4PV$) calculated from thermal pressure
acting as a reliable estimate of energy content,
when in actual fact the CR and magnetic energy densities 
may dominate \citep[e.g.][]{mathews_creation_2008}. 

Based on the UHECRs arriving at Earth with energies 
above $55$~EeV and directions 
clustered around Centaurus A, 
\cite{joshi_very_2018} estimate an UHECR luminosity of $\sim10^{39}$~erg~s$^{-1}$. The jet powers 
in Centaurus A, Fornax A and M87 exceed this value by orders of magnitude.
However, it seems that the current jet powers in 
these sources struggle, similarly to starbust winds, to meet
the power requirements (equation~\ref{eq:power})
for particle acceleration to high 
energy. 
Despite this, the {\em average} powers in radio galaxies
can be greater than $P_{\mathrm{min}}$, and the peak powers still greater,
suggesting that past jet activity is important.

\subsection{Enhanced activity in the past?}
Based on the current energetics and distances
alone, we might expect M87 to contribute a similar 
UHECR flux to Fornax A, but there is not a clear hotspot 
close to M87 in the observed UHECR data. 
However, the jet powers in local radio galaxies 
could feasibly have been different in the past. 
Acceleration during a more
luminous phase aids with power requirements and allows DSA to
operate at fast shocks, which is important since 
the shocks associated with the currently
active Centaurus A jet struggle to accelerate the highest energy CRs 
\citep{croston_high-energy_2009}. 
There is evidence in Centaurus A
and Fornax A for enhanced activity within the last 
$\sim100$~Myr. Both show giant lobes with 
linear sizes greater than $250$~kpc,
whose energy contents are large compared to the current 
power of the jet; $\sim5\times10^{58}$~erg in Fornax A 
\citep{lanz_constraining_2010} and as high as 
$10^{59-60}$~erg in Centaurus A \citep{wykes_mass_2013,eilek_dynamic_2014}.
The energy content of the lobes in M87 is lower,
approximately $8\times10^{57}$~erg 
\citep{mathews_creation_2008} and the 
lobes only extend $\sim80$~kpc across
\citep{owen_m87_2000}. The M87 lobes are generally 
consistent with being inflated 
by the current jet \citep{owen_m87_2000}, 
whereas Fornax A and Centaurus A hint at a more violent past.

In Centaurus A, \cite{wykes_mass_2013} estimate the 
buoyancy time at $\tau_{\mathrm{buoy}} = 560$~Myr, which 
places a lower limit
on the jet power to inflate the giant lobes of 
$5\times10^{43}$~erg~s$^{-1}$. 
This jet power could feasibly have been much higher. 
Fornax A shows direct evidence of declining AGN activity 
\citep{iyomoto_declined_1998,lanz_constraining_2010},
and {\em both} sources are thought to have undergone
mergers \citep{mackie_evolution_1998,horellou_atomic_2001}, 
with Fornax A showing evidence for merger activity 
within 3~Gyr \citep{goudfrooij_kinematics_2001,goudfrooij_star_2001},
and potentially as recently as $0.1$~Gyr \citep{mackie_evolution_1998}. 
Mergers can trigger AGN activity 
as they provide fuel that can subsequently accrete onto 
a central black hole
\citep[e.g.][]{blundell_inevitable_1999,hopkins_cosmological_2008,silverman_impact_2011}.

Both Centaurus A and Fornax A seem promising candidates for a 
scenario in which a merger-triggered
AGN outburst produced more powerful jets in the past,
accelerating UHECRs that are still escaping from the giant
lobe reservoirs. The Larmor radius of an UHECR proton is 
\begin{equation}
r_g \approx 
\frac{E}{10~\mathrm{EeV}} 
\left( \frac{B}{10\mu\mathrm{G}} \right)^{-1}~\mathrm{kpc},
\end{equation}
indicating that long-term containment in 
the much-larger $100$ kpc-scale lobes is likely.
The magnetic field lines advected with the jet 
material that ultimately produces the 
lobes do not connect to the ambient medium. 
UHECRs are confined to local magnetic field lines, 
so UHECR escape requires the crossing of field lines, which is a slow process.
\citep[e.g.][]{ozturk_trajectories_2012,zweibel_microphysics_2013}. 
It is therefore not safe to assume
that the UHECRs are accelerated in the present source state;
it is the history of the source over the shorter of the GZK time and the UHECR escape time that matters.  This also means that the energy content of the lobes could make a good estimate of UHECR luminosity since it is an integrated measure over past activity.
The sound-crossing time -- the timescale 
for adiabatic losses -- in Centaurus A is on the order of 
$\tau_{\mathrm{buoy}}$ \citep{wykes_mass_2013}.
This is longer than the GZK time of
$r_{\mathrm{GZK}}/c\approx300$~Myr, 
which implies that adiabatic losses are unimportant. 
GZK and hadronic $\gamma$-ray losses might still matter
for $\gamma$-ray emission.
In fact, UHECRs in the Centaurus A lobes are thought to produce
some of the observed $\gamma$-ray flux 
\citep{sahu_hadronic-origin_2012,joshi_very_2018}.
Overall, it seems reasonable that the UHECRs are  
escaping on a timescale roughly comparable to the time since 
the outburst ended, but shorter than a GZK time.

It has been suggested that the UHECRs can also be accelerated by an 
in-situ, ongoing process in the lobes, such as second-order Fermi 
\citep{fraschetti_ultra-high-energy_2008,hardcastle_high-energy_2009}.
However, \cite{hardcastle_which_2010} notes that
this would require relativistic turbulence to reach the 
required energies. 

\section{Conclusions}
We have shown that the observed excesses in the UHECR arrival 
directions measured with the 
PAO is more naturally explained by association 
with radio galaxies than with SBGs. 
Although SBGs are favoured in the A18 analysis,
we argue that the increased significance they report compared with 
their $\gamma$AGN sample is largely driven by one source 
near the south Galactic pole (NGC 253) and an increased flux estimate in the 
vicinity of Centaurus A due to the nearby SBGs M83 and NGC4945. 
If Centaurus A were more luminous, as indeed it
is in the 3FHL catalogue or if the lobes contribution is accounted for, 
and Fornax A were included, then 
this can increase the significance of 
the $\gamma$AGN result, provided that we allow for 
reasonable magnetic deflection of around $20^\circ$ and  
decreased attenuation compared to A18 scenario A.

We suggest that radio galaxies are likely 
candidates for UHECR production. 
Building on previous work
\citep[e.g.][]{norman_origin_1995,romero_centaurus_1996,rachen_ultra-high_2008,fraschetti_ultra-high-energy_2008,eichmann_ultra-high-energy_2018}, 
we have introduced a physical scenario to account for
UHECR production in fading giant radio lobes from a recent 
jet ``outburst''. 
This scenario could be further developed to 
apply to SBGs with past AGN activity; radio galaxies and SBGs 
need not be unrelated populations. Further work from PAO 
coupled with more detailed modelling work should help 
to discriminate further. 

\section*{Acknowledgements}
We are grateful to the anonymous referee for a constructive 
and helpful report. We would like to 
thank Jonathan Biteau and Alan Watson 
for helpful correspondence
regarding magnetic deflection and the PAO analysis.
We would also like to thank 
Garrett Cotter, Robert Laing and Justin Bray for 
useful discussions. 
This work is supported by the Science and Technology 
Facilities Council under grants ST/K00106X/1 and ST/N000919/1. A.T.A. thanks the Czech Science Foundation (ref. 14-37086G) -  ``Albert 
Einstein Center for Gravitation and Astrophysics'' in Prague, and the 
EU COST Action (ref. CA16104) 
``Gravitational waves, black holes and fundamental physics''.
We acknowledge the use of matplotlib \citep{matplotlib}
and astropy \citep{the_astropy_collaboration_astropy_2018}.

\vspace{-1em}
\bibliographystyle{mnras}
\bibliography{fornax,software} 
\bsp	
\label{lastpage}
\end{document}